\documentclass[english,aps,prd,a4paper,groupedaddress,preprintnumbers,floatfix,nofootinbib,showpacs]{revtex4}

\usepackage{graphicx}
\usepackage{enumerate}

\usepackage{amsmath}
\usepackage{amssymb}
\usepackage{amsfonts}
\usepackage{xcolor}
\usepackage{url}
\usepackage{hyperref}
\usepackage{booktabs}
\usepackage{caption}

\hypersetup{colorlinks=true,linkcolor=redLinks,citecolor=greenLinks,
urlcolor=redLinks,
pdfborder={0 0 1}}
\hypersetup{
    bookmarks=true,         % show bookmarks bar?
    unicode=false,          % non-Latin characters in Acrobat?s bookmarks
    pdftoolbar=true,        % show Acrobat?s toolbar?
    pdfmenubar=true,        % show Acrobat?s menu?
    pdffitwindow=false,     % window fit to page when opened
    pdfstartview={FitH},    % fits the width of the page to the window
    pdftitle={My title},    % title
    pdfauthor={Author},     % author
    pdfsubject={Subject},   % subject of the document
    pdfcreator={Creator},   % creator of the document
    pdfproducer={Producer}, % producer of the document
    pdfkeywords={keyword1, key2, key3}, % list of keywords
    pdfnewwindow=true,      % links in new PDF window
    colorlinks=false,       % false: boxed links; true: colored links
    linkcolor=red,          % color of internal links (change box color with linkbordercolor)
    citecolor=green,        % color of links to bibliography
    filecolor=magenta,      % color of file links
    urlcolor=cyan           % color of external links
}

\newcommand{\bi}{\begin{itemize}}
\newcommand{\ei}{\end{itemize}}

\begin{document}
\title{The reactor antineutrino anomaly and low energy threshold neutrino experiments}
\author{B. C. Ca\~nas$^{1,2}$}
\email{blanca.canas00@usc.edu.co} 
\author{E. A. Garc\'es$^1$}
\email{egarces@fis.cinvestav.mx} 
\author{O. G. Miranda$^1$}
\email{omr@fis.cinvestav.mx} 
\author{A. Parada$^2$}
\email{alexander.parada00@usc.edu.co} 
\affiliation{$^1$~Departamento de F\'isica, Centro de Investigaci\'on y de Estudios Avanzados 
del IPN, Apdo. Postal 14-740, 07000 Ciudad de M\'exico, M\'exico.}
\affiliation{$^2$~Universidad Santiago de Cali, Campus Pampalinda, Calle
5 No. 6200, 760001, Santiago de Cali, Colombia}
%%%%%%%%%%%%%%%%%%%%%%%%

\begin{abstract}
Short distance reactor antineutrino experiments measure an
antineutrino spectrum a few percent lower than expected from theoretical
predictions.  
In this work we study the potential of low energy threshold reactor
experiments in the context of a light sterile neutrino
signal. We discuss the perspectives of the recently
  detected coherent elastic neutrino-nucleus scattering in  future
  reactor antineutrino experiments. We find that the
  expectations to improve the current constraints on the mixing with
  sterile neutrinos 
  are promising.
We also analyse the measurements of antineutrino scattering off
electrons from short distance reactor experiments. In
  this case, the statistics is not competitive with inverse beta decay
  experiments, although future experiments might play a role when
  compare it with the Gallium anomaly.
\end{abstract}


\maketitle
%%%%******************************* Introduction ****************************
\section{Introduction}\label{sec-intro}
%%%%***************************** SECCION 2 *****************************
Neutrino physics is already in the precision physics era; with
  recent Nobel prize awarded in 2015 and with most of the Standard
  Model parameters already measured with good
  accuracy~\cite{deSalas:2017kay,Esteban:2016qun,Capozzi:2016rtj}. Future neutrino experiments will
  try to improve the determination of these parameters, especially the
  neutrino CP violating phase~\cite{Acciarri:2015uup}.
  Besides oscillations, there is also a complete program of neutrino
  experiments aiming to improve the measurements of neutrino cross
  sections~\cite{Aliaga:2013uqz,Drakoulakos:2004gn}. 

Historically, reactor neutrino experiments have been a powerful tool in
the measurement of neutrino electron scattering~\cite{Reines:1976pv}. 
Recently, several experiments have
measured this process with an increased precision~\cite{Vidyakin:1992nf,Derbin:1993wy,Amsler:1997pn,Deniz:2009mu}
and it is expected that new results will be reported in the near
future, for instance by the GEMMA experiment~\cite{Beda:2012zz}. 
Despite the small cross section, neutrino electron scattering data
have given 
interesting results on neutrino properties, such as neutrino
magnetic moments~\cite{Canas:2015yoa}, as well as on the value of
the weak mixing angle at low energies~\cite{Canas:2016vxp}.

Regarding inverse beta decay (IBD) experiments, besides the successful
measurements of the standard oscillation parameters, both for
long~\cite{Eguchi:2002dm,Abe:2008aa} and for short
baselines~\cite{An:2016ses,RENO:2015ksa,Abe:2014bwa}, there is also a
complete program to unambiguously discover or exclude sterile
neutrinos in the near future.  Some of these experiments are underway
and others will start data taking
soon~\cite{Boireau:2015dda,Ashenfelter:2015uxt,Serebrov:2016wzv}. The
DANNS experiment~\cite{Alekseev:2017iqo} has already presented
preliminary results. On the other hand, recent results from the NEOS
experiment already exclude part of the previously allowed region in
the most recent 3+1 sterile neutrino data fit~\cite{Ko:2016owz}.

Also in the low energy threshold regime, there is the coherent elastic
neutrino nucleus scattering (CENNS), that was studied for the first
time in the seventies~\cite{Freedman:1973yd} and has finally been
observed~\cite{Akimov:2017ade}. A large number of proposals are also
looking for this signal, and there will be several measurements of the
neutrino cross sections with this reaction in the future. As it has been
proved by the COHERENT Collaboration~\cite{Akimov:2017ade}, CENNS is a
very promising process for low energy neutrino physics. Several works
have pointed out its impact in testing non-standard
interactions~\cite{Barranco:2005yy,Scholberg:2005qs,Lindner:2016wff,Coloma:2017egw,Barranco:2011wx,Shoemaker:2017lzs},
neutrino magnetic
moment, or the
 weak mixing
angle~\cite{Wong:2005pa,Kosmas:2015vsa,Kosmas:2015sqa}. 

Recently, the
sensitivity of CENNS to a sterile neutrino has been studied for the
case of the Texono and the COHERENT
proposals~\cite{Kosmas:2017zbh}. 
Since the revaluation of the reactor antineutrino energy
spectrum~\cite{Mueller:2011nm}, the possibility of an additional
sterile neutrino~\cite{Mention:2011rk} has been under scrutiny.  Most
of the evidence for this anomaly comes from short baseline reactor
experiments using IBD and from the so-called Gallium
anomaly~\cite{Laveder:2007zz,Giunti:2010zu}.
In this work we also study the case of a light sterile neutrino,
considering a wider set of experimental proposals that plan to use
CENNS.  We focus in the case of reactor antineutrino fluxes. In this
sense, our work compares different proposals that use a similar
antineutrino flux and discuss the advantages and complementarities of
these future experiments.  

At the same time, we also discuss in more
detail the case of a different prescription for the reactor
antineutrino flux as a solution to the so called reactor anomaly.
After the recent evaluation of the antineutrino spectrum by Daya
Bay~\cite{An:2015nua}, the need for a better understanding of the
spectrum has been pointed out.  Moreover, the possibility that the
reactor anomaly can be solved by a revaluation of the antineutrino
flux has also been considered~\cite{Giunti:2016elf}. 
Since the data in the reactor signal for sterile neutrinos come from
IBD experiments, it will be interesting to consider alternative
detection technologies as a complementary test to this anomaly. For
this reason we study here the current data from neutrino electron
scattering, as well as the prospects of CENNS.

%%%%%%%%%%%%%%%%%%%%%%%%%%%%%%%%%%%%%%%%%%%
\section{Antineutrino electron scattering measurement}
%%%%%%%%%%%%%%%%%%%%%%%%%%%%%%%%%%%%%%%%%%

In this section we concentrate our study in experiments that use the
electron antineutrino scattering off electrons as the detection
process. For this purpose, we have reanalyzed the experimental
results, using the current prescription for the reactor antineutrino
flux~\cite{Mueller:2011nm}, to obtain a restriction on the mixing
parameters of a sterile neutrino. Following this approach, the
effective survival probability for short baseline antineutrino
experiments in the 3+1 mixing scheme~\footnote{A different oscillation
  channel to a sterile neutrino would be that of a $\nu_\mu\to\nu_s$
  transition, as hinted by the LSND~\cite{Athanassopoulos:1997pv} and
  MiniBooNE~\cite{Aguilar-Arevalo:2013pmq} Collaborations. Since we
  focus in a different channel, for this case we refer the reader to
  the limits reported in
  Refs~\cite{Adamson:2016jku,Aartsen:2017bap,Adamson:2017zcg}.} can be
written as ~\cite{Abazajian:2012ys}
\begin{equation}\label{eq:survprob}
 P_{{\bar\nu}_{e}\rightarrow{\bar\nu}_{e}}^\text{SBL}
=  \sin^{2}2\theta_{ee} 
\sin^{2}\left(\frac{\Delta m_{41}^{2}L}{4E}\right),
\end{equation}
where
\begin{equation}
 \sin^{2}2\theta_{ee} = 4|U_{e4}|^{2}(1-|U_{e4}|^{2}).
\end{equation}

\noindent The expected number of events, in the presence of 
a fourth, sterile, neutrino state, will be given in this case as 
\begin{equation}
N_{i} = n_e\Delta t \int \int^{T_{i+1}}_{T_i}\int  
                   \lambda(E_{\nu})  P_{\nu_{\alpha}\rightarrow\nu_{\alpha}}^{\text{SBL}} 
                   \frac{d\sigma}{dT}  
                  R(T,T') dT' dT dE.
\label{diff:cross:sec-sm}
\end{equation}
where $\lambda(E_{\nu})$ stands for the antineutrino spectrum; for
energies above $2$~MeV, this spectrum has been taken according to
Ref~\cite{Mueller:2011nm}; on the other hand, if we need to include
energies bellow $2$~MeV, we have included the spectrum computed in
Ref.~\cite{Kopeikin:1997}.  $R(T,T')$ is the resolution function for
the given experiment,
$P_{\nu_{\alpha}\rightarrow\nu_{\alpha}}^{\text{SBL}}$ is the effective survival
probability as given in Eq.~(\ref{eq:survprob}), and
$\frac{d\sigma}{dT}$ is the differential
cross section for the antineutrino scattering off electrons,
given as~\cite{Vogel:1989iv}
\begin{equation}
\frac{d\sigma}{dT} = \frac{2G^2_Fm_e}{\pi}
              \left[g^2_R + g^2_L (1-\frac{T}{E_\nu})^2 - g_Lg_R m_e\frac{T}{E_\nu^2}\right] ,
\label{eq:difxs}
\end{equation}
where $m_e$ stands for the electron mass and $G_F$ is the Fermi
constant. In this expression, $g_L = 1/2 + \sin^2\theta_{\rm W}$ and $g_R
= \sin^2\theta_{\rm W}$ are the usual
Standard Model couplings.

Several experiments using neutrino electron scattering as detection
reaction have been performed along the years. Some of them have
searched for a non-zero neutrino magnetic
moment~\cite{Giunti:2014ixa}. The experiments for our analysis will be TEXONO, MUNU, Rovno and Krasnoyarsk. The most recent experimental result has been
given by the TEXONO Collaboration~\cite{Deniz:2009mu}, that has
reported the measurement of ten bins with an electron recoil energy
between $3$ and $8$~MeV. The energy resolution for this experiment was
$\sigma(T)=0.0325 \sqrt{T}$~\cite{Wong:2006nx}. A previous experiment,
with a lower threshold, was performed by the MUNU
Collaboration~\cite{Daraktchieva:2005kn}. In this case, the error in
the electron recoil energy was considered to be $\sigma(T)=0.08
\text{T}^{0.7}$~\cite{Daraktchieva:2003dr}. We also considered the
Rovno~\cite{Derbin:1993wy} and Krasnoyarsk~\cite{Vidyakin:1992nf}
results. For these experiments, the fuel proportions, as well as the
electron recoil energy window, are shown in Table~\ref{table:exps}.

\begin{figure}
\includegraphics[width=0.45\textwidth]{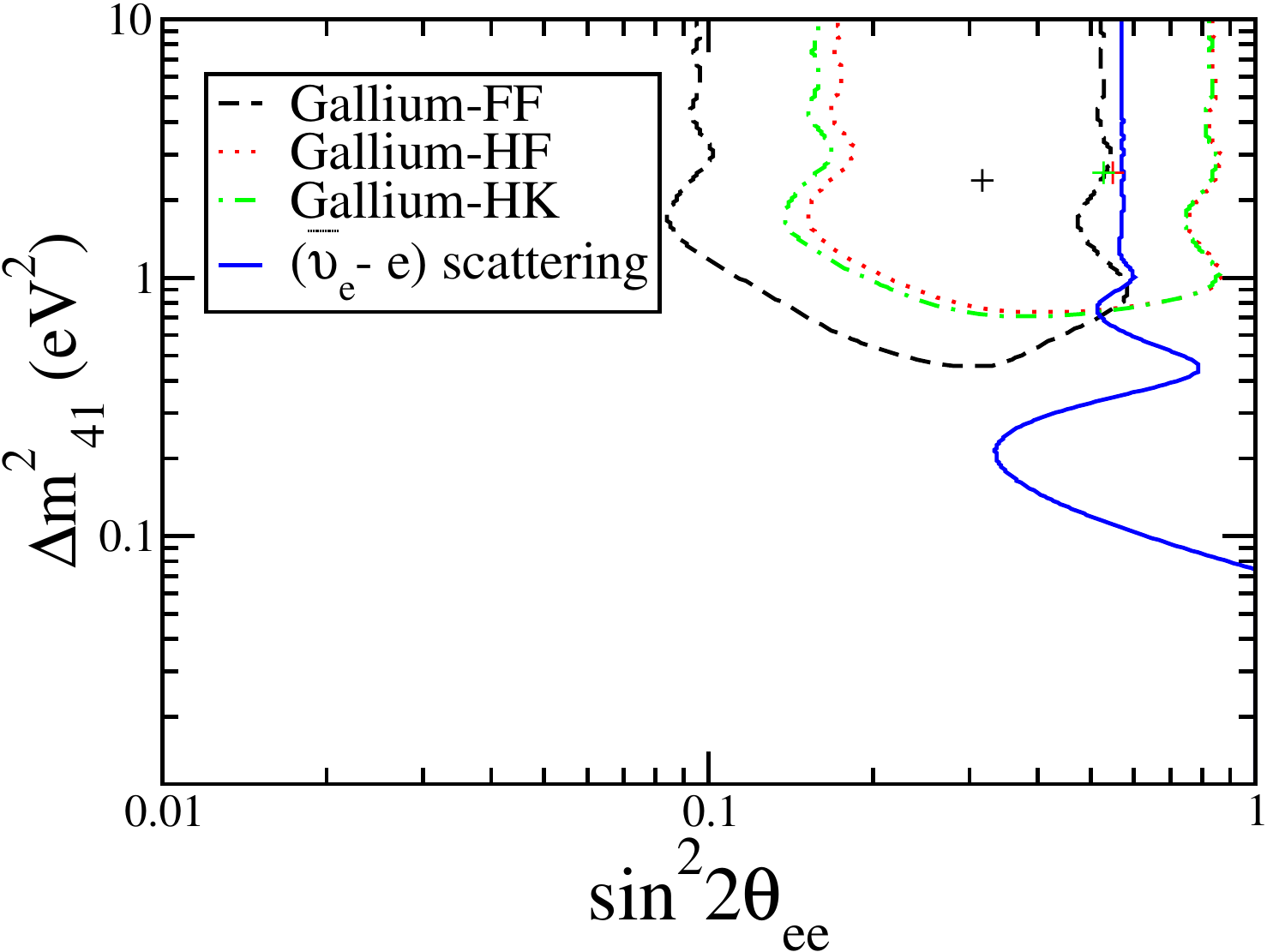}
\caption{\label{fig:3} Restrictions for a sterile neutrino  from a
  combined analysis of neutrino electron scattering from reactor
  experiments at 90 \% C L (blue solid line). We also show for comparison, the 
  results for the
  Gallium anomaly~\cite{Giunti:2012tn} in the three cases discussed in the text.
The best-fit values 
are indicated by a cross. }
\end{figure}

\begin{widetext}
\begin{table}[!t]
\begin{center}
\begin{tabular}{l c c c c c c c c c c c} \hline \hline
Experiment               &  &  $^{235}$U   &      & $^{239}$Pu      &  & 
$^{238}$U    &    $^{241}$Pu & & $T_{thres}$ & & observable\\ \hline \hline
TEXONO~\cite{Deniz:2009mu}    &     & $0.55$      &   &  
$0.32$     & &  $0.07$  &  $0.06$  & & $3-8$~MeV & & $\sigma = (1.08\pm0.21\pm0.16)\cdot \sigma_{SM}$\\ 
MUNU~\cite{Amsler:1997pn}     &     & $0.54$      & &  
$0.33$     & & $0.07$     &   $0.06$   &  &  $0.7-2$~MeV & & $(1.07\pm0.34)$~events/day \\
Rovno~\cite{Derbin:1993wy}       &     & $\simeq1.0$      &   &  
$-$     & &  $-$  &  $-$  & & $0.6-2$~MeV  
& & $\sigma = (1.26\pm 0.62)\times10^{-44}$cm$^2$/fission\\
Krasnoyarsk~\cite{Vidyakin:1992nf} &     & $\simeq1.0$      &   &  
$-$     & &  $-$  &  $-$    & & $3.15-5.175$~MeV 
& & $\sigma = (4.5\pm 2.4)\times10^{-46}$cm$^2$/fission\\ \hline \hline 
\end{tabular}
\caption{Summary of the measured $\bar{\nu}_{e}-e$ scattering cross
  sections from reactor antineutrino experiments. The columns show the
  fuel averaged proportions, the electron recoil energy window, and
  the reported observables.}
\label{table:exps}
\end{center}
\end{table}
\end{widetext}

 We have performed a goodness of fit analysis for the experiments
 quoted above.  After performing the combined fit using the four
 reactor experiments, we have obtained the restriction for the sterile
 oscillation parameters, $\sin^{2}2\theta_{ee}$ and $\Delta
 m_{41}^{2}$, as shown in Fig.~(\ref{fig:3}).  We also show in this
 figure the allowed regions for the Gallium
 anomaly~\cite{Laveder:2007zz,Giunti:2010zu}.  We have followed the
 procedure described by Giunti et
 al.~\cite{Giunti:2010zu,Giunti:2012tn}, with the only difference of
 including in our analysis the updated results for the Gamow-Teller
 transitions reported by Frekers {\it et al.}~\cite{Frekers:2015wga}
 (FF). This result is shown as a dash line in Fig.~(\ref{fig:3}).
 This recent measurement has also been considered for the
  case of the future experiment BEST~\cite{Barinov:2017ymq}. As it is
  possible to notice, the current resolution from electron
  antineutrino scattering off electrons has no overlap with this
  region. Therefore, the constraint obtained here would be of interest
  only if one considers other measurements of the Gamow-Teller
  transitions, such as that of in the $(p,n)$ experiment of Krofcheck
  {\it et al.}~\cite{Krofcheck:1985fg} (HK, dash-dotted line) or the shell
  model of Haxton~\cite{Haxton:1998uc} (HF, dotted line), also
  considered in the Ref~\cite{Giunti:2012tn}. It would be expected
  that new measurements of the antineutrino-electron scattering could
  be more restrictive, as in the case of the proposed GEMMA updated
  experiment~\cite{Beda:2012zz}.  Still, despite the increased
  interest in solving the Gallium anomaly~\cite{Barinov:2017ymq},
  current global analysis on the sterile
  signal~\cite{Gariazzo:2017fdh,Dentler:2017tkw} give a region that is
  in tension with the large value of $\sin^22\theta_{ee}$ obtained
  from the Gallium data. For that reason we discuss in the next
  section the case of coherent elastic neutrino nucleus scattering as
  a promising technique to give complementary information to that
  coming from inverse beta decay experiments.  

\section{Perspectives for Coherent neutrino nucleus scattering in reactor experiments}

\noindent The CENNS is another interesting process to explore physics
beyond the Standard Model. This interaction was proposed more than
four decades ago within the SM
context~\cite{Freedman:1973yd,Drukier:1983gj}. Different
Collaborations and experimental proposals have considered the
possibility of detecting the coherent neutrino-nucleus
scattering~\cite{Wong:2008vk,Aguilar-Arevalo:2016qen,Dutta:2015vwa,Collar:2014lya}. Recently
the COHERENT Collaboration has achieved the first detection of CENNS,
opening a promising new era of low energy neutrino experiments.

In this section we will study four different proposals that plan to
use a reactor as their antineutrino source. They are the TEXONO,
MINER, RED100, and CONNIE experiments, that we describe briefly in
what follows.

\begin{itemize} 
\item The TEXONO Collaboration has proposed the use of high purity
  Germanium-based detectors, with a threshold energy of $T_{thres}\sim
  100$~eV~\cite{Wong:2008vk,Soma:2014zgm}. The Collaboration expects to 
  develop a modular detector and reach $1$~kg mass for the target. 
  The reactor flux would come from 
  the Kuo-Sheng nuclear power plant and the detector would be located
  $28$~m away from the reactor. For a quenching factor $Q_f=0.25$ the expected 
  number of events would be $4000$~kg~$^{-1}$~year$^{-1}$~\cite{Wong:2008vk}. 
\item The MINER Collaboration will use a detector made of $^{72}$~Ge
  and $^{28}$~Si with a $2:1$ proportion and with a threshold energy, 
   $T_{thres}\sim 10$~eV.  A TRIGA-type pool reactor
  will deliver an antineutrino flux with a fuel average proportion
  of ($^{235}$~U:$^{238}$~U:$^{239}$~Pu:$^{241}$~Pu) given
  by~\cite{Dutta:2015vwa} ($0.967$:$0.013$:$0.02$:$0.001$). With this
  special type of reactor, the detector can be located at a distance
  of $1-3$~m from the source. An event rate of
  $5-20$~kg~$^{-1}$~day$^{-1}$ is forecast for this
  configuration~\cite{Agnolet:2016zir}. In our simulations we will
  consider a $20$~kg $^{72}$~Ge detector with one year of data taking at 
  an event rate of $5$~events~kg~$^{-1}$~day$^{-1}$. 
\item The Kalinin power plant has also a program to detect CENNS. At
  least two different options appear in the literature. One is a
  germanium detector, $\nu$GeN~\cite{Belov:2015ufh}, while the other
  one considers the use of liquid Xenon, RED100~\cite{Akimov:2012aya}.
  We focus in the Xenon case as this material has been of interest for
  different experimental groups~\cite{Baxter:2017ozv} and it is a
  different target with an energy threshold of $T_{thres}\sim
  0.5$~keV~\cite{Santos:2011ju}. The expected distance to the Kalinin
  reactor is about $19$~m and they expect to detect $1020$ events per
  day~\cite{Akimov:2017hee}.  The expected fiducial mass is
  $100$~kg~\cite{Akimov:2012aya}. As in the previous proposals, we
  consider one year of data taking.

\item The CONNIE Collaboration~\cite{Aguilar-Arevalo:2016qen} is
  currently working at the Angra-2 reactor using Charged-Coupled
  Devices (CCD's) as a detector, at $30$~m from the
  reactor. It is expected that CCD technology can reach
    an energy threshold of $28$~eV~\cite{Moroni:2014wia} and
    $16.1$~events per day for one kg of material. We will consider
    again one year of data taking. 
\end{itemize}

In order to calculate the number of events for any of the above proposals, we
use the following expression for the cross section,

\begin{equation} 
\left(\frac{d\sigma}{dT}\right)_{\rm SM}^{\rm coh} = \frac{G_{F}^{2}M}{2\pi}\left[1-\frac{MT}{E_{\nu}^{2}}+\left(1-\frac{T}{E_{\nu}}\right)^{2}\right]\left\{
 [(Zg^{p}_{V}+Ng^{n}_{V})F(q^2))]^{2}\right\}
\end{equation}

\noindent here, $M$ is the mass of the nucleus, $E_{\nu}$ is the neutrino
energy, $T$ is the nucleus recoil energy, $F(q^{2})$ is the nuclear
form factor, and the neutral current vector couplings (including
radiative corrections) are given by~\cite{Barranco:2005yy}
\begin{eqnarray}
\nonumber g_{V}^{p} &=&\rho_{\nu
  N}^{NC}\left(\frac{1}{2}-2\hat{\kappa}_{\nu
  N}\hat{s}_{Z}^{2}\right)+2\lambda^{uL}+2\lambda^{uR}+\lambda^{dL}+\lambda^{dR}\\ g_{V}^{n}
&=&-\frac{1}{2}\rho_{\nu
  N}^{NC}+\lambda^{uL}+\lambda^{uR}+2\lambda^{dL}+2\lambda^{dR}
\end{eqnarray}
\noindent where $\rho_{\nu N}^{NC}=1.0082$,
$\hat{s}_{Z}^{2}=\sin^2\theta_{W}=0.23126$, $\hat{\kappa}_{\nu
  N}=0.9972$, $\lambda^{uL}=-0.0031$, $\lambda^{dL}=-0.0025$, and
$\lambda^{dR}=2\lambda^{uR}=7.5 \times 10^{-5}$~\cite{Beringer:1900zz}. 
We have checked that, for a first analysis of the expected sensitivity
to a sterile neutrino signal, the corresponding form factors,
$F(q^{2})$, will not play a significant role\footnote{We have computed
  the form factor with the effective model of Ref.\cite{Engel:1991wq}.}
and, therefore we have taken them as unity in what follows.
For estimating the number of expected events (SM) in the detector, we
use the expression,
\begin{equation}\label{n_events}
 N^{\rm SM}_{\rm events}=t\phi_{0}\frac{M_{\text{detector}}}{M}\int_{E_{\nu \rm min}}^{E_{\nu \rm max}}\lambda(E_{\nu})dE_{\nu}\int_{T_{\rm min}}^{T_{\rm max}(E_{\nu})}\left(\frac{d\sigma}
 {dT}\right)_{\rm SM}^{\rm coh} dT ,
\end{equation}
\noindent where $M_{\text{detector}}$ is the mass of the detector,
$\phi_{0}$ is the total neutrino flux, $t$ is the data taking time
period, $\lambda(E_{\nu})$ is the neutrino spectrum, $E_{\nu}$ is the
neutrino energy, and $T$ is the nucleus recoil energy.  The maximum
recoil energy is related with the neutrino energy and the nucleus mass
through the relation $T_{\rm
  max}(E_{\nu})=2E_{\nu}^{2}/(M+2E_{\nu})$. In all the cases we will
consider one year of data taking.

For the oscillation to a fourth sterile family, we will consider the
two families case in vacuum, where the number of events is 

\begin{equation}\label{n_events_NS}
 N^{\rm NS}_{\rm events}=t\phi_{0}\frac{M_{\text{detector}}}{M}\int_{E_{\nu \rm min}}^{E_{\nu \rm max}}\lambda(E_{\nu})P_{\nu_{\alpha}\rightarrow\nu_{\alpha}}^{\text{SBL}} 
dE_{\nu}\int_{T_{\rm min}}^{T_{\rm max}(E_{\nu})}\left(\frac{d\sigma}
 {dT}\right)_{\rm SM}^{\rm coh} dT .
\end{equation}

In the above equation,
$P_{\nu_{\alpha}\rightarrow\nu_{\alpha}}^{\text{SBL}}$ represents the neutrino survival probability as expressed in
Eq.~(\ref{eq:survprob}). The differential cross section has just been
discussed above, and the antineutrino flux will depend on the specific
reactor under consideration. With this expression we can make a
forecast for different experimental setups. We will consider the case
of the MINER, RED100, and TEXONO proposal with the fluxes and
thresholds mentioned above. We will assume that each
experiment will measure exactly the standard prediction for the  
three active neutrino picture. With this hypothesis we will obtain an 
expected $\chi^2$ analysis assuming only statistical errors. 

\begin{figure}[h!] 
        \includegraphics[width=0.45\linewidth]{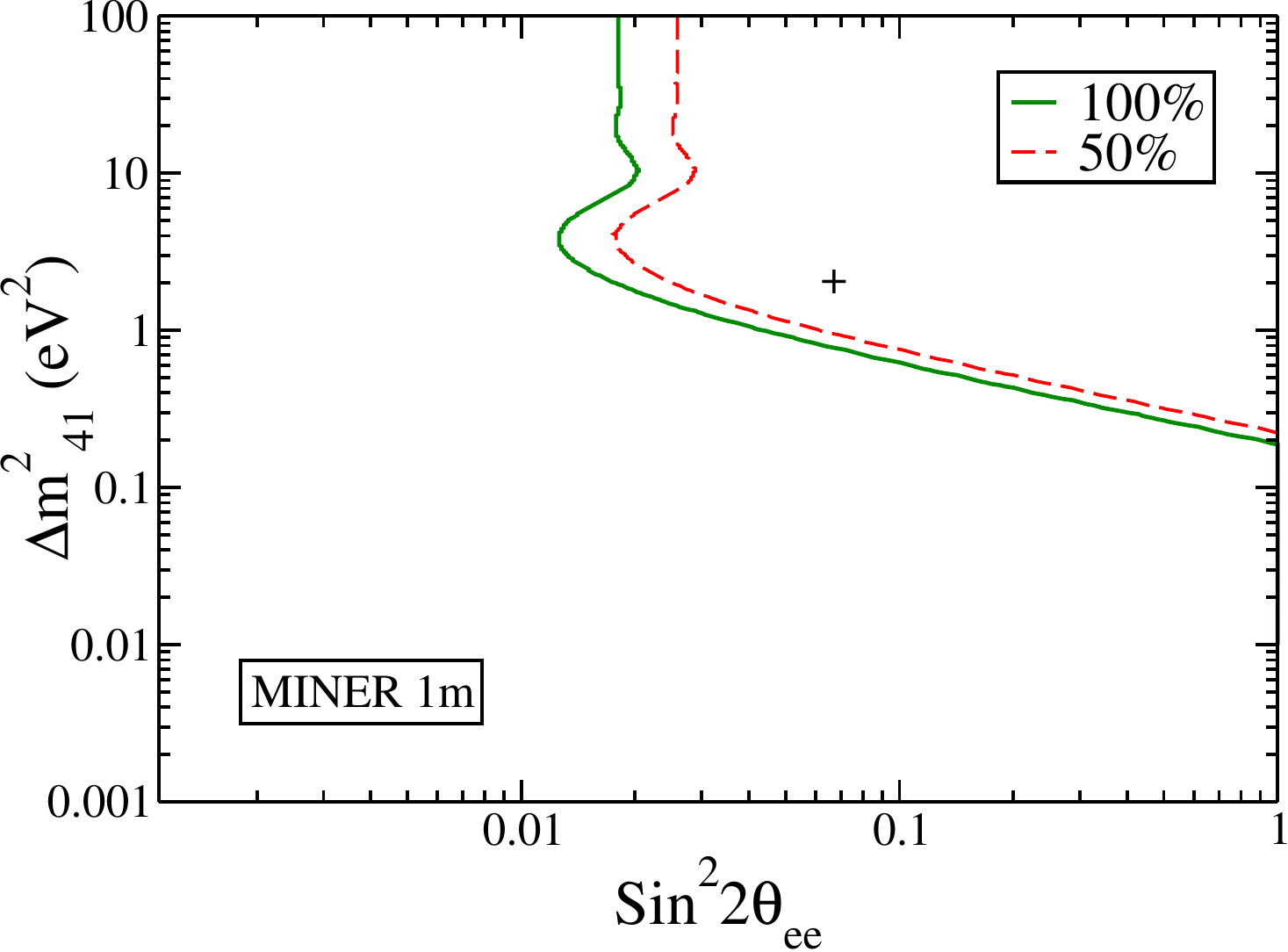}
        \includegraphics[width=0.45\linewidth]{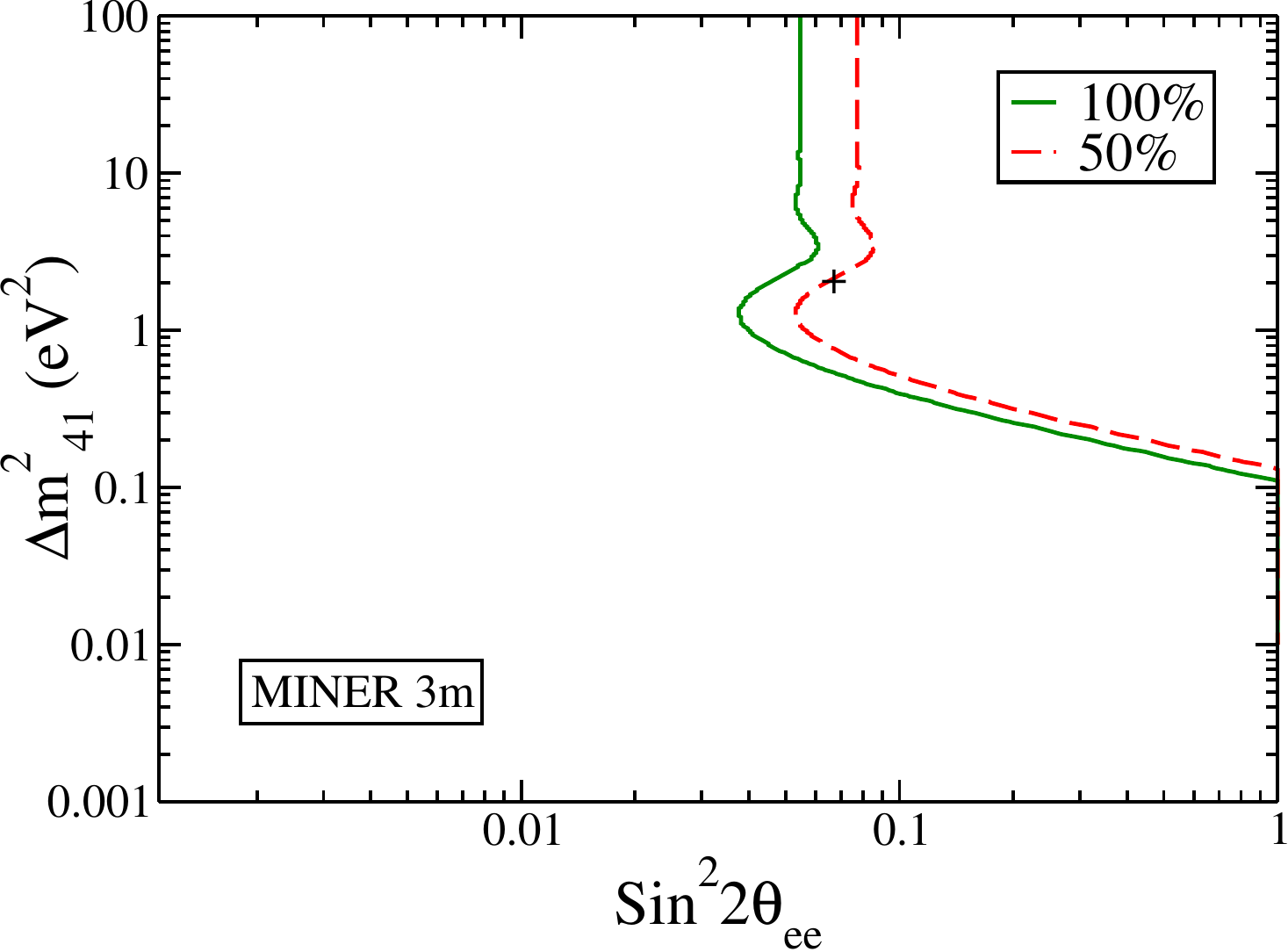} 
    \caption{\label{fig:miner} Exclusion regions for the MINER
      experiment. The left (right) panel corresponds to a baseline of
      $1$ ($3$)~m.  The solid (green) line is for a detector with
      100\% efficiency and the dashed (red) line is for a 50\%
      efficiency. The current best fit point for the sterile analysis
      is shown as a reference. }
\end{figure}

\begin{figure}[h!]
        \includegraphics[width=0.45\linewidth]{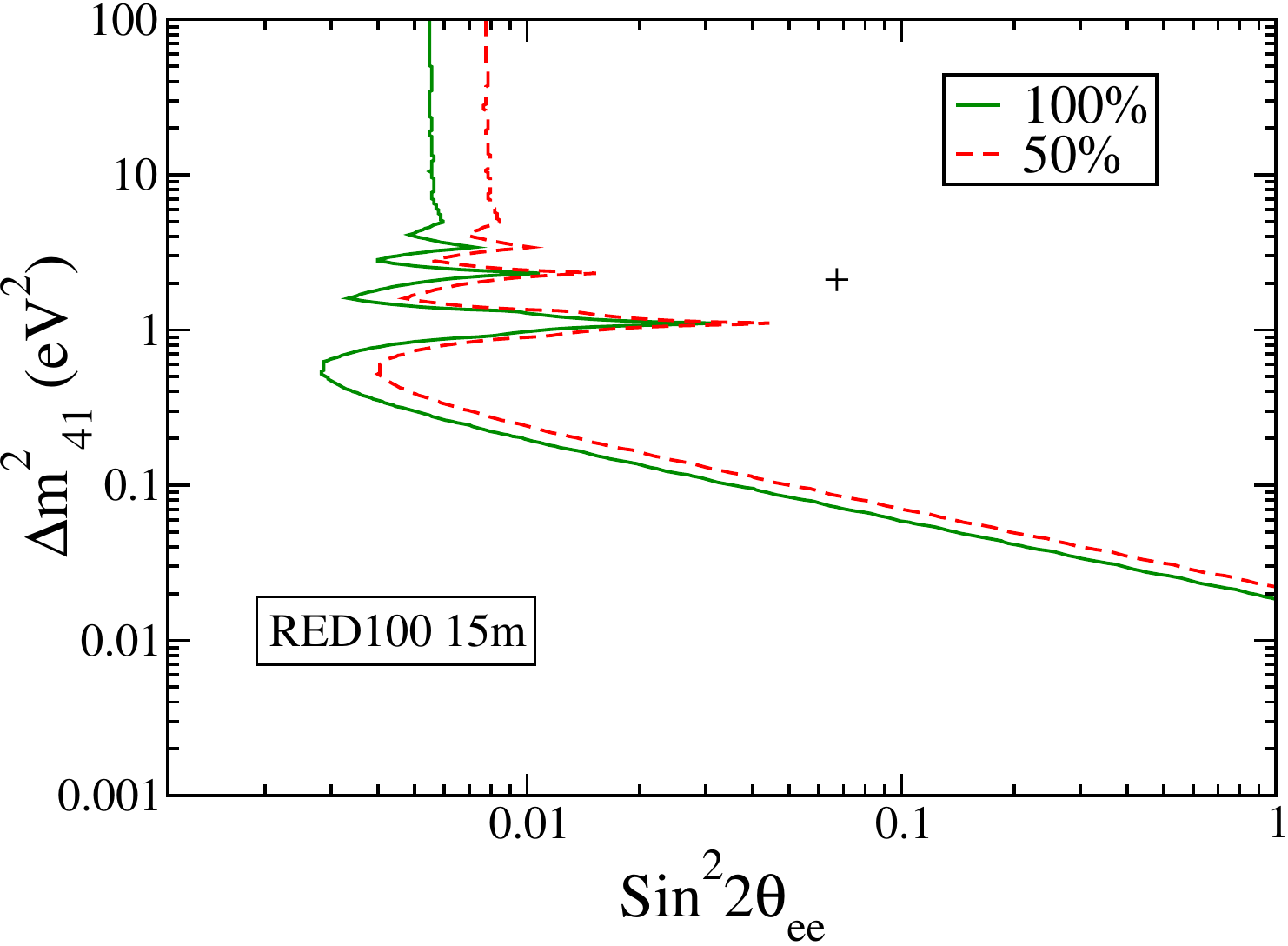} 
        \includegraphics[width=0.45\linewidth]{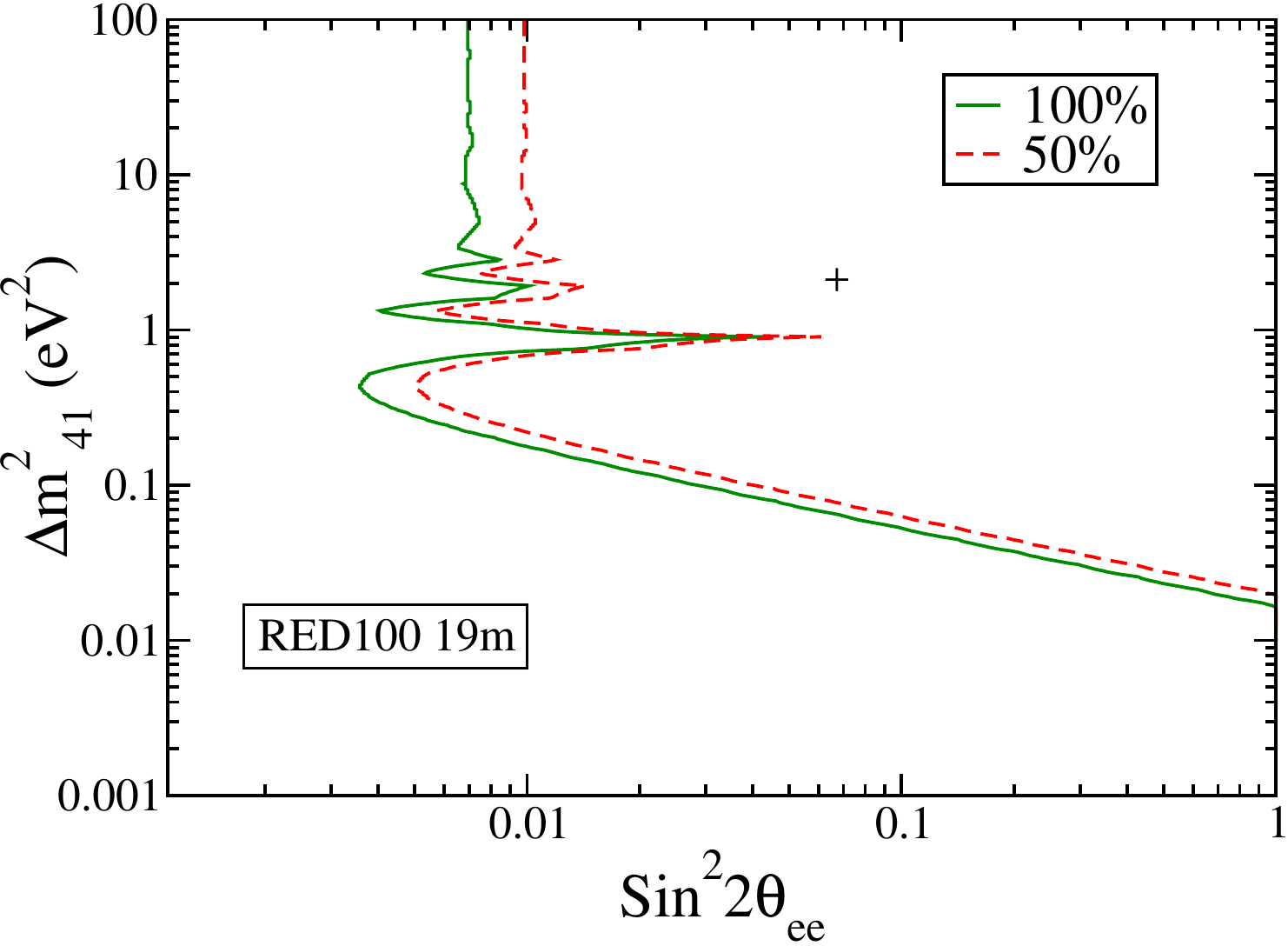} 
    \caption{\label{fig:red100} Exclusion regions for the RED100
      proposal. The left (right) panel shows the case of a baseline of 
      $15$ ($19$)~m. The solid (green) line
      correspond to a detector with 100\% efficiency and the dashed
      (red) lines  a 50\% efficiency. The current best fit point for the 
      sterile analysis is shown as a reference. }
\end{figure}

\begin{figure}[h!]
  \includegraphics[width=0.5\textwidth]{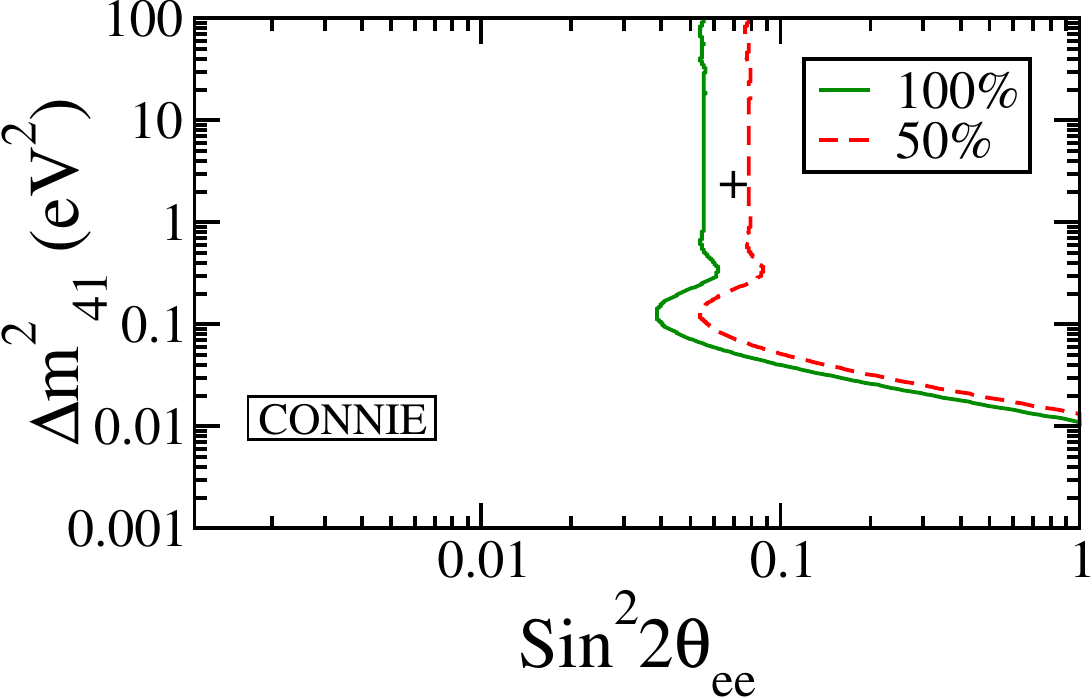}
    \caption{\label{fig:CONNIE} Exclusion regions for the CONNIE 
      proposal. 
      The solid (green) line
      correspond to a detector with 100\% efficiency and the dashed
      (red) lines  a 50\% efficiency. 
       The current best fit point for the 
      sterile analysis is shown as a reference. }
\end{figure}

The result of these computations for the MINER Collaboration is shown
in the Fig.~(\ref{fig:miner}), where we have considered two different
baselines of $1$~m and $3$~m. Since we are using only statistical
errors, our analysis can be considered as very optimistic. In order to
consider the more realistic counterpart, we have also shown in the
same figure the case where the detector can only achieve a $50$~\%
efficiency. We can notice that for a baseline of $1$ m the MINER
Collaboration could exclude the current best fit point to the sterile
neutrino analysis~\cite{Gariazzo:2017fdh}. A similar analysis was done
for the case of the RED100 proposal where we have considered the
Kalinin nuclear power plant as the antineutrino flux source. We show
in Fig.~(\ref{fig:red100}) the case of two different baselines and two
possible efficiencies. The expectative to improve the current
constraints on the mixing with a sterile neutrino is even more
promising in this case, despite the relatively high detection energy
threshold. In the case of the CONNIE proposal, we have
  performed a similar analysis, shown in Fig.~(\ref{fig:CONNIE}) where
  it is also possible to reach the region of interest for the sterile
  signal.
\begin{figure}%[t] 
\begin{center}
\includegraphics[width=0.5\textwidth]{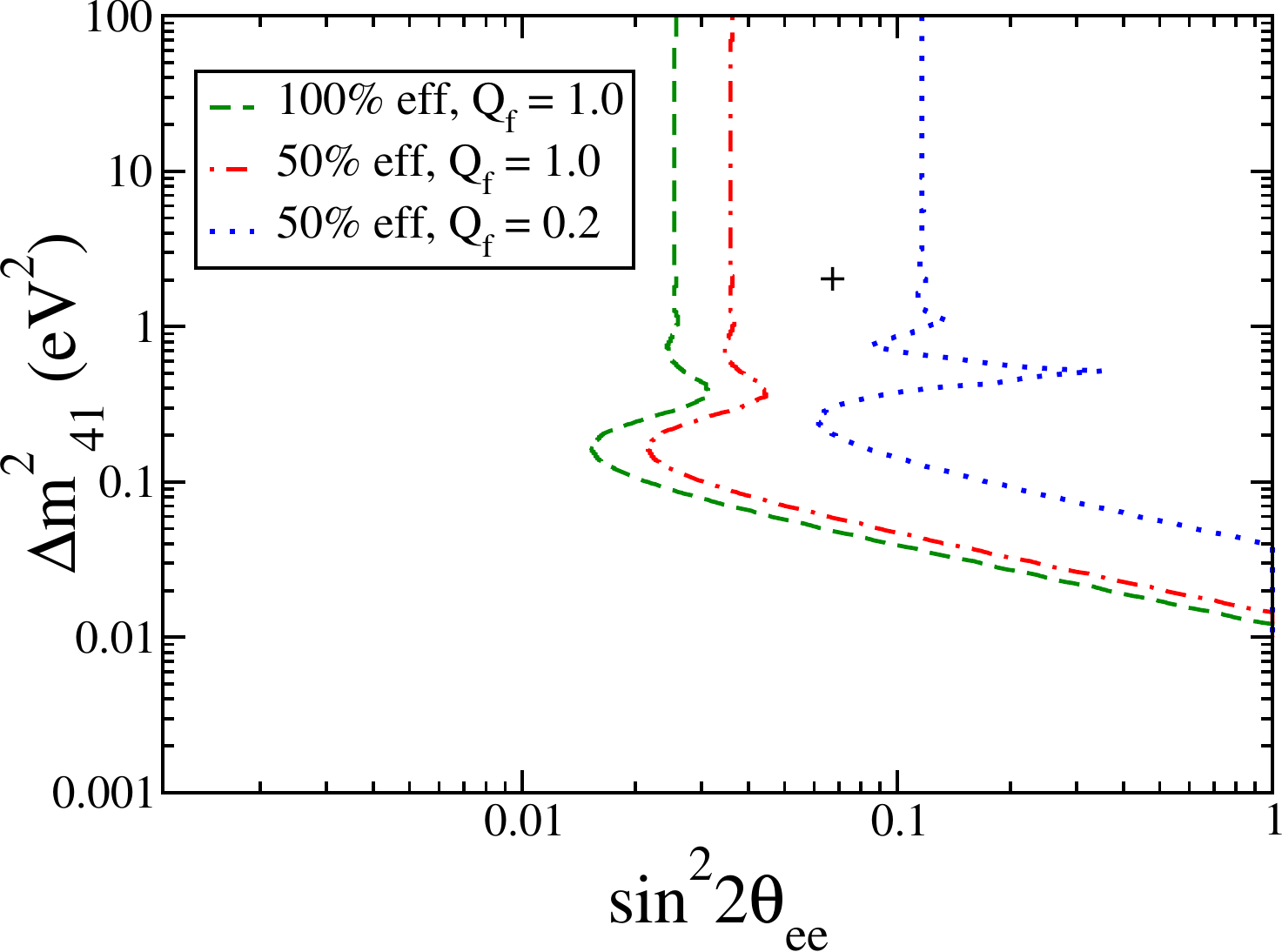}
\end{center}
\caption{\label{fig:texono}Expected sensitivity for a reactor
  antineutrino experiment with a detector based on CENNS. We consider
  the TEXONO proposal as a reference with a Germanium detector. The
  three different exclusion regions correspond to a 90 \% CL for
  different combinations of efficiency and energy threshold: The most
  restrictive region is for a $100$~\% efficiency and a $100$~eV
  threshold ($Q_f=1$), while the less restrictive case is for a
  $50$~\% efficiency and a relatively high energy threshold of
  $500$~eV ($Q_f=0.2$). Finally the intermediate case shown in the
  figure is for a $50$~\% efficiency and a $100$~eV threshold
  ($Q_f=1$).}
\end{figure}
\begin{figure}%[ht] 
\begin{center}
\includegraphics[width=0.55\textwidth]{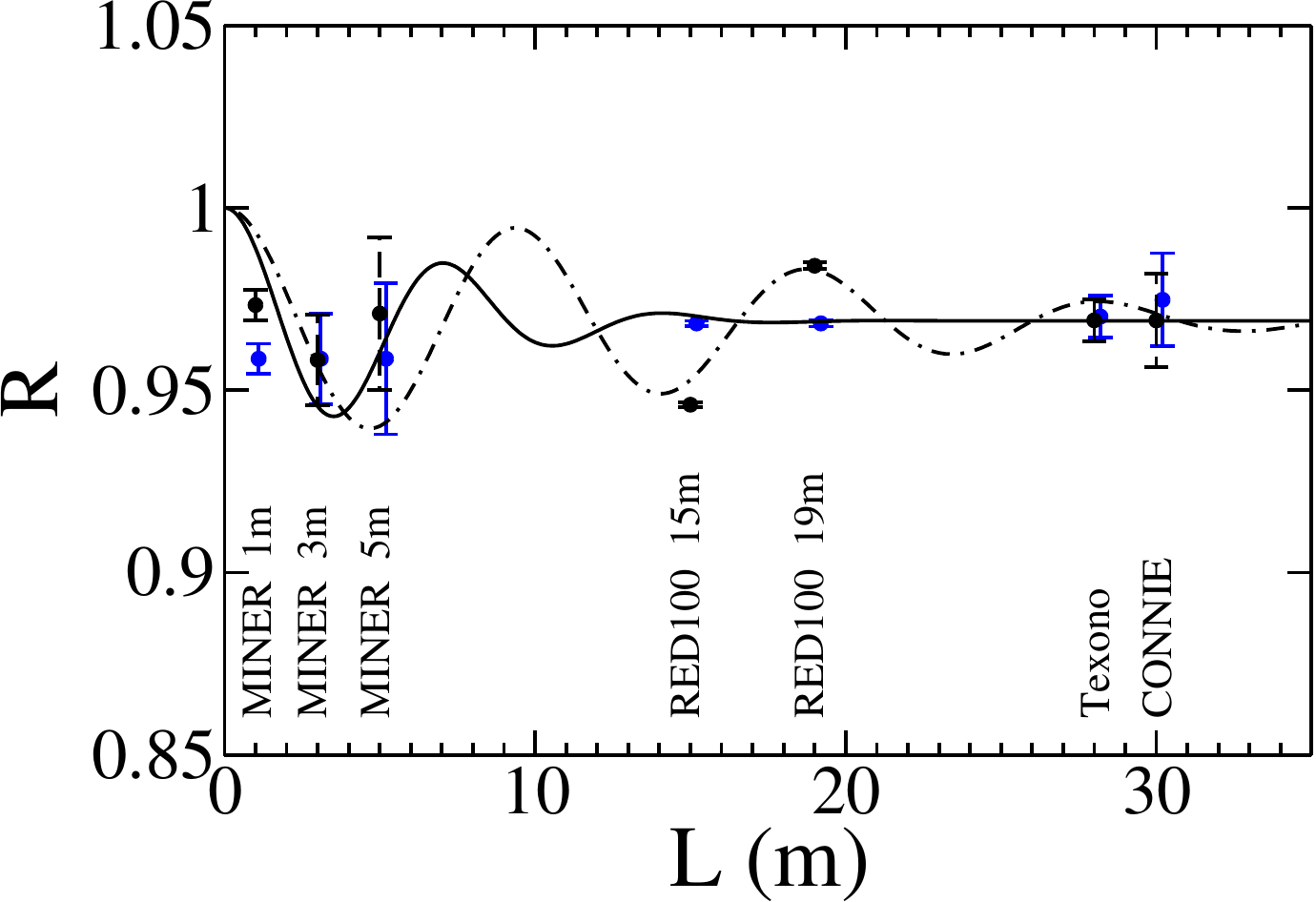}
\end{center}
\caption{\label{fig:cnnall} Ratios $R$ of predicted to expected rates
  for different proposed CENNS experiments. We have taken the Mueller 
  spectrum as a reference in our calculations. Different baselines are shown for  some detectors, taking into account that the proposals are still under 
  discussion. The black dots show the
  expected ratio for the case of a sterile neutrino with a
  $\sin^22\theta_{ee}=0.062$ and $\Delta m^2 = 1.7$~eV. The blue dots
  give the ratio for the case of a decrease in the $^{235}$U of $5$~\% as proposed in a recent article~\cite{Giunti:2016elf}. The black line 
  represents the average probability for a mean
  energy of $4$~MeV, and the dotted black curve corresponds to an energy of 
  $6.5$~MeV, both with an energy resolution of $15$~\%. And finally the error bars account for the statistical errors.}
\end{figure}

\noindent We have also analyzed the case of the TEXONO proposal. The
results are shown in Fig.~(\ref{fig:texono}).  As in the previous
cases, we have also considered different possibilities for this
proposal. In particular, we take into account different quenching
factors for the detector. This factor represents the ratio of the
electron recoil to nucleus recoil energy~\cite{Simon:2002cw}, which
gives us an important correction since the detector response to a
nucleus recoil energy is different from the response coming from
electron calibration sources.  The quenching factor is given by
\begin{equation}
Q_{f} = \frac{E_{ee}}{E_{Nr}},
\end{equation}
\noindent where $E_{ee}$ represents the electron equivalent energy and
$E_{Nr}$ is the nuclear recoil energy.  In the case of the TEXONO
experiment, we calculated the expected number of events for the
quenching factors $Q_{f}=1$~and $Q_{f}=0.2$.

The regions of mixing angle and squared-mass splitting favored by
different combinations of quenching factors and detector efficiencies
are shown in the Fig.(\ref{fig:texono}). The results are in agreement
with the previous work of Ref.\cite{Kosmas:2015vsa} and shows other
cases with a different quenching factor.  The expectations for this
proposal are competitive with the MINER and RED100 proposals as can be
seen from Figs.~(\ref{fig:miner}) and (\ref{fig:red100}).

We conclude this section comparing the expected signal for these
proposals in two very different situations.  Recently, the theoretical
estimates for the antineutrino flux have been under deep scrutiny
(see for
  instance~\cite{Giunti:2016elf,An:2017osx,Giunti:2017yid}) and the
  reactor anomaly might be solved by a re-evaluation of the neutrino
  fluxes~\cite{Giunti:2016elf,Giunti:2017yid}. In this case, it is
also possible that the CENNS experiments give a confirmation of this
result, especially if several CENNS experiments with different
baselines are performed, as seems to be the case. This situation is
illustrated in Fig.~(\ref{fig:cnnall}), where we show what will be the
antineutrino rate measured by these proposals if a $5$~\% decrease in
the $^{235}$~U is considered~\cite{Giunti:2016elf} (without any
sterile effect).
On the other hand, we also show the expected
ratio for the same experiments, in the case that the sterile neutrino
is the responsible for the deficit. For this case we consider $\Delta
m^2 = 1.7$~eV~$^2$ and $\sin^22\theta_{ee} = 0.062$, according
to the most recent fit of antineutrino disappearance data~\cite{Gariazzo:2017fdh}.  
As expected, the different baselines will
give a different ratio for the sterile solution. 
The situation is different if the reactor anomaly is due to a
correction in the antineutrino flux, where the expected number of
events will be different than for the oscillation explanation,
especially for the RED100 and the MINER ($1$~m) cases.
In this case, as expected, the complementarity of different
experiments using different baselines, thresholds, and  fuel
proportions could be very helpful in discriminating what is the real
explanation of the reactor anomaly.

\section{Conclusions}

In this work we have studied the reactor anomaly in the context of
future CENNS experiments and in antineutrino electron scattering data
from short baseline reactor neutrino experiments.
Concerning antineutrino-electron scattering we conclude that this
interaction can give limited information due to the relatively poor
statistics.
On other hand, the recent observation of CENNS 
by the COHERENT Collaboration strongly motivates the further
exploration of physics beyond the Standard Model in this context.  
We show that CENNS experiments could play an important role in the
determination, or exclusion, of the sterile signal.  Particularly, the
RED100, TEXONO, MINER, and CONNIE proposals could test the current best fit
point of the sterile allowed parameter space.
Regarding the need of a precise antineutrino flux determination, CENNS
is particularly attractive, since the detection technique is different
from that of IBD detectors.  In this case, we obtained the ratios
between predicted and expected data in two different cases:
considering sterile neutrinos and taking a decrease in the
antineutrino flux as it is suggested by some recent works. Both
situations could be of interest in order to explain the reactor
antineutrino anomaly.

\acknowledgments \noindent We thank Dmitri Akimov and Alexis
Aguilar-Arevalo for useful discussions. This work has been supported
by CONACyT. E. A. G. thanks the CONACYT Project No. FOINS-296-2016
(Fronteras de la Ciencia).  A. Parada was supported by
  Universidad Santiago de Cali (USC) under grant
  DGI-COCEIN-No.935-621717-016. B. C. Ca\~nas is also supported by USC
  through the DGI.\\\\

%\end{references} 

%
\end{document}